\documentclass[11pt]{article}

\date{}

\usepackage{url}
\usepackage{cite}
\usepackage{plain}

\usepackage{wrapfig}
\usepackage{graphics}
\usepackage{picins,graphicx}
\usepackage[english]{babel}
\usepackage[leqno]{amsmath}
\usepackage{amssymb}
\usepackage{verbatim}
\usepackage[mathscr]{eucal}

\usepackage{amstext}
\usepackage{makeidx}
\usepackage{amsthm}

\usepackage{hyperref}
\hypersetup{colorlinks,%
citecolor=red,%
linkcolor=blue,%
urlcolor=black}

\makeindex

\newtheorem{theorem}{Theorem} 

\newtheorem{proposition}{Proprieta'}
\newtheorem{definition}{Definizione}
\newtheorem{notation}{Nota}
\newtheorem{ex}{Esercizio} 
\newtheorem{esempio}{Esempio}

\newcommand{\vs}{\vspace{3mm}}
 
\newcommand{\no}{\noindent} 

\newcommand{\beq}{\begin{equation}} 
\newcommand{\eeq}{\end{equation}}

\newcommand{\bex}{\begin{ex}} 
\newcommand{\eex}{\end{ex}} 

\newcommand{\bese}{\begin{esempio}} 
\newcommand{\eese}{\end{esempio}} 

\newcommand{\bpro}{\begin{proposition}} 
\newcommand{\epro}{\end{proposition}}

\newcommand{\bthe}{\begin{theorem}} 
\newcommand{\ethe}{\end{theorem}}

\newcommand{\bnote}{\begin{notation}} 
\newcommand{\enote}{\end{notation}}

\newcommand{\bdefi}{\begin{definition}} 
\newcommand{\edefi}{\end{definition}} 

\newcommand{\bc}{\begin{center}} 
\newcommand{\ec}{\end{center}}

\newcommand{\mail}[1]{\href{unina:#1}{\texttt{#1}}}

\usepackage{pdfsync}

\author{Monica De Angelis\thanks{Univ. of Naples  "Federico II", Dip. Mat. Appl. "R.Caccioppoli", \newline
 Via Claudio n.21, 80125, Naples, Italy.
\newline\mail{modeange@unina.it}}}

\title {On a parabolic operator of dissipative systems}

\begin{document}
\maketitle

\begin{abstract}
\no A parabolic integro differential operator operator $\mathcal L$   	suitable to describe many   phenomena in various physical fields,is considered. By means of equivalence between   $\mathcal L$  and the third order  equation which describe the evolution inside an exponentially shaped Josephson junction (ESJJ),  an  asymptotic analysis  for (ESJJ) is achieved, evaluating explicitly   boundary contributions related to  the Dirichlet problem.

\vs \no {\bf{Keywords}}:{ Reaction - diffusion systems;\hspace{2mm} Biological applications;\hspace{2mm} Laplace  transform,\hspace{2mm}FitzHugh Nagumo model.}

\vs \no \textbf{Mathematics Subject Classification (2000)}\hspace{1mm}35E05  \hspace{1mm}35K35\hspace{1mm} 35K57 \hspace{1mm}35Q53 \hspace{1mm} 78A70
\end{abstract}


\vspace{13mm}


\section{Introduction }

\setcounter{equation}{0}
\setcounter{theorem}{0}

The present paper deals with a third order differential equation which  characterizes  a lot of physical models linear or not. Many  analogies   among  nonlinear  operators and p.d.e. systems exist and the bibliography is large. (see, f.i. \cite{rb,acscott,acscott02}). Here,  the  semilinear equation, which characterizes exponentially shaped Josephson junctions (ESJJ) in superconductivity, is considered \cite{mda13,df13,df213,ddf}  and  equivalence  with  the following  parabolic integro differential equation:  
\beq   \label{11}
  \mathcal L u \equiv \,\, u_t -  \varepsilon  u_{xx} + au +b \int^t_0  e^{- \beta (t-\tau)}\, u(x,\tau) \, d\tau \,=\, F(x,t,u) \,
   \eeq  
  
  \no  is proved. In this way   a priori estimates are obtained  and, by means of of a well known  
theorem  on asymptotic behavior of convolutions, 
  asymptotic effects of boundary perturbations for the solutions of  Dirichlet and Neumann boundary value problems are achieved.

   Operator  $ {\cal L}  $  defined in  (\ref{12}) can  describe  many physical phenomena and there is plenty  of bibliography.\cite{bcf,r,dr1,mps,fr,l} Moreover,  assuming   $ \varepsilon, a, b, \beta  $  as  positive constants,  previous analyses have been done in  \cite{dm13,de13,dr13,dr8}. The fundamental solution $ K $ has been determined and many its properties have been proved. Moreover,  Neumann, Dirichlet and mixed boundary value problems have been considered and non linear integral equations, whose Green  functions have numerous   properties typical of the diffusion equation, have been determined. When $F = F(x, t, u)$,  equation (\ref{12})  involves some non linear phenomena both in superconductivity and in biology where  characterizes  reaction diffusion models like the FitzHugh-Nagumo system suitable to  model the propagation  of nerve impulses.
   
   In superconductivity it is well known that the Josephson effect   is modeled by the perturbed sine Gordon equation (PSGE) given by (sse,f.i. \cite{bp}):
\begin{equation}  \label{12}
\varepsilon u_{xxt}\, - \, u_{tt} \, +\, u_{xx}-\, \alpha u_t = \,  \, \sin u \ + \gamma   
\end{equation} 

\no where $u$ represents the difference between the phase of the wave functions for the two superconductors of the junction, $ \gamma$ is a forcing term that is proportional to a bias current, the $\alpha$-term accounts for dissipation from normal electrons crossing the junction, and the 
$\varepsilon$-term accounts for dissipation from normal electrons flowing parallel to the junction.

Besides, when the case of an exponentially shaped Josephson junction (ESJJ) 
 is considered, denoting by  $ \lambda$  a positive constant,  the evolution of the phase inside the junction is described by the third order equation:

 \beq            \label{13}
  \varepsilon \varphi_{xxt}+\varphi_{xx}-\varphi_{tt} - \varepsilon\lambda \varphi_{xt} -\lambda  \varphi_x- \alpha\varphi_{t}=\sin
  \varphi-\gamma
 \eeq
 \no where terms  $ \,\,\lambda \,\varphi_x $ and $\,\lambda \, \varepsilon \,\varphi_{xt} \,$ represent the current due to the tapering and in particular $ \lambda\varphi _{x} $ corresponds to a geometrical force driving the fluxons from the wide edge to the narrow edge. \cite{bcs00,cmc02,bcs96}. An  exponentially shaped Josephson junction  provides several advantages with respect to a rectangular junction. For instance it is possible to obtain  a voltage which is not  chaotic anymore, but rather periodic  excluding, in this way,   some among the possible causes of large spectral width. It is also proved that  the problem of trapped flux can be avoided \cite{bss,j05,j005,ssb04}. Moreover, some devices as SQUIDs were built with exponentially tapered loop areas.\cite{cwa08}
 
   According  to  the meaning of source term $ F(x,t,u) $  equation (\ref{11}) describes the
evolution of several  models. Besides, the  kernel  $ e^{- \beta (t-\tau)}\, u(x,\tau) $   can be modified   as physical situations demand. Particular choose done here is  due to superconductive  and biological models considered. Indeed,I have already said,it it possible to prove that since equation (\ref{11})both  the FitzHugh Nagumo system  and PSGE  (\ref{12}) can be deduced. Moreover, as proved further,  the evolution in an  Josephson junction  with nonuniform width can be characterized,too.

   \section{Statement of the problem}
   
      \setcounter{equation}{0}
\setcounter{theorem}{0}

Denoting by   $\, T\, $    an arbitrary positive constant let  

\[
\,   \Omega_T \, \equiv \{\,(x,t) : \, 0\,\leq \,x \,\leq L \,\,;  \ 0 < t \leq T \,\}. \]

  Let us consider the following initial boundary value problem with  Dirichlet boundary conditions assigned to equation (\ref{11}):

\vspace{3mm}
\begin{equation}   \label{21}
\left \{
   \begin{array}{lll}
{  \cal L}\,u\,  \,=\, F(x,t,u) \, & (x,t) \in \Omega_T \,  \\    
\\  \,u (x,0)\, = u_0(x)\, \,\,\, &
x\, \in [0,L], 
\\
\\
  \, u(0,t)\,=\,\varphi_1(t)  \qquad u(L,t)\,=\,\varphi_2(t) & 0<t\leq T.
   \end{array}
  \right.
\end{equation}

Denoting by   $ \, K(x,t) \, $ the fundamental solution of the linear  operator defined by (\ref{11}),one has \cite {dr8}:

\beq     \label {22}
K(r,t)=  \frac{1}{2 \sqrt{\pi  \varepsilon } }\,\biggl[ \frac{ e^{- \frac{r^2 }{4 t}-a\,t\,}}{\sqrt t}
 -\,b \int^t_0  \frac{e^{- \frac{r^2}{4 y}\,- a\,y}}{\sqrt{t-y}}  \,\ e^{-\beta ( t \,-y\,)}  J_1 (2 \sqrt{\,by(t-y)\,}\,)dy \biggr]
\eeq

\vs\no where $\, r\,= |x| \, / \sqrt \varepsilon \, \, $ and   $ J_n (z) \,$    denotes the Bessel function of first kind and order $\, n.\,$ One has:

\vspace{3mm}\begin{theorem}

For all $t>0$, the Laplace transform of  $\,K (r,t)\, \,$  with respect to $\,t\,$ converges absolutely in the half-plane $ \Re e  \,s > \,max(\,-\,a ,\,-\beta\,)\,$ and it results:

\vs
\beq      \label{23}
\,\hat K\,(r,s)  =\,\int_ 0^\infty e^{-st} \,\, K\,(r,t) \,\,dt \,\,=  \, \frac{e^{- \,r\,\sigma}}{2 \, \sqrt\varepsilon \,\sigma \,  }
\eeq

\vs \no with $\,\sigma^2 \ \,=\, s\, +\, a \, + \, \frac{b}{s+\beta}.$ 
\end{theorem}

Let us  now consider the following  Laplace transforms with respect to $\,t\,$:

\vs
\[
\hat u (x,s) \, = \int_ 0^\infty \, e^{-st} \, u(x,t) \,dt \,\,, \,\,\,  \,\hat F (x,s)   \, = \int_ 0^\infty \,\, e^{-st} \,\, F\,[x,t,u (x,t)\,] \,dt \,,\
\]

\vs \no  and let $\hat \varphi_1(s ), \,\,\, \hat \varphi_2(s)\,\, $  be the  ${ L} $  transforms of the  data $ \varphi_i(t ) \,\,(i=1,2).\, $

\vs \noindent  Then the Laplace transform of the problem (\ref{21})  is formally given by:

\beq   \label{24}
\left \{
   \begin{array}{lll}
  \hat u_{xx}  \,\,- \dfrac{\sigma^2}{\varepsilon} \,\,\hat u =\, -\,\dfrac{1}{\varepsilon} \,\,[\, \,\hat F(x,s,\hat u(x,s)) +u_0(x)\,\,]\\    
\\
  \,\hat u_x(0,s)\,=\, \hat \varphi_1\,(s)\qquad \hat u_x(L,s)\,=\,\hat \varphi_2\,(s).
   \end{array}
  \right.
\eeq

\vspace{3mm}\noindent   If one introduces the following {\it theta function }

\beq\,  \label{25}
\begin{split}
\displaystyle
\hat \theta \,(\,y,\sigma)\,= 
\\ & \frac{1}{2 \,\, \sqrt\varepsilon \,\,\,\sigma  } \, \biggl\{\, e^{- \frac{y}{\sqrt \varepsilon} \,\,\sigma}+\, \sum_{n=1}^\infty \,\, \biggl[ \,e^{- \frac{2nL+y}{\sqrt \varepsilon} \,\,\sigma} \, +\, e^{- \frac{2nL-y}{\sqrt \varepsilon} \,\,\sigma}\,
\biggr] \, \biggr\}
\\ \\& =\dfrac{\cosh\,[\, \sigma/\sqrt{\varepsilon} \,\,(L-y)\,]}{\,2\, \, \sqrt{\varepsilon} \,\, \sigma\,\,\, \sinh\, (\,\sigma/\sqrt{\varepsilon}\,\, \,L\,)}\,\,=
\\&
\end{split}\eeq

\vspace{3mm}\noindent then, by  (\ref{24}) and (\ref{25}) one deduces:

\begin{equation}     \label{26}
\hat u (x,s) = \,\int _0^L \, [\,\hat \theta\,(\,x+\xi, \,s\,)\,-\,\,\,\hat \theta\,(\,|x-\xi|,\, s\,)\,] \, \,[\,u_0(\,\xi\,) \,+\,\hat f(\,\xi,s)\,]\,d\xi\, \end{equation}
\\
\[
\displaystyle -\, 2 \,\,\varepsilon \, \,\hat g_1 \,(s) \,\,   \frac{\partial}{\partial \,x}\,\, \hat\theta (x,s)\,+ \, 2 \,\, \varepsilon  \,\, \hat g_2 \, (s)\,\,\frac{\partial}{\partial \,x}\,\,\hat  \theta \,(L-x,s\,).\, \,\,\]

\section{Explicit solution}

So, in order to obtain the inverse formula for (\ref{26}),         let us  apply (\ref{23}) to (\ref{25}). Then,  one deduces the  following function which is  similar to  {\em theta functions}:

\begin{equation}     \label{37}
\theta (x,t) \,=\,  K_0(x,t) \ +\, \sum_{n=1}^\infty \,\, \ [\, K_0(x \,+2nL,\,t) \, + \, K_0 ( x-2nL, \,t)\,] \, =  
\end{equation}
\[\, =\sum_{n=-\infty }^\infty \,\, \ K_0(x \,+2nL,\,t). \,\]

  As  a consequence, when $\, F\, =\, f(x,t), \,$ by (\ref{25}) the explicit solutions of  the {\em linear} problem $\, (P_D )\,$  is:

\begin{equation}   \label{389}
 u(\, x,\,t\,)\, = \,\,\int^L_0 \, [\theta \,(x+\xi,\, t)\,- \theta (|x-\xi|,\,t)\,]\, \,u_0(\xi)\,\, d\xi \,\,+ \,
\end{equation}

\[ - \,2 \,  \varepsilon \,\int^t_0  \frac{\partial}{\partial \,x}\,\,\theta\, (x,\, t-\tau) \,\,\, g_1 (\tau )\,\,d\tau\,+\, 2\,\, \varepsilon \int^t_0 \frac{\partial}{\partial \,x}\,\,\theta\, (L-x,\, t-\tau) \,\,\, g_2 (\tau )\,\,d\tau\,\]

\[ +\, \,\int^t_0 d\tau\int^L_0 \, [\,\theta\, (x+\xi,\, t-\tau)- \theta (|x-\xi|,\,t-\tau )] \,\,\, f\,(\,\xi,\tau\,)\, \,\,d\xi.\]

 Owing to the basic properties of $ K_0(x,t), $ it is easy to deduce the following theorem:

  \begin{theorem}
When the linear source $ \, f(x,t)\,  $ is continuous in $ \Omega_T\, $ and the initial boundary  data $ u_0(x),$ $ g_i\,\,(i=1,2)$  are continuous, then  problem $(P_D) $ admits a unique regular solution $ u(x,t)  $ given by (\ref{389}).
\end{theorem}

\section{Equivalence with  (ESJJ)}    
   
   \setcounter{equation}{0}
\setcounter{theorem}{0}

Among many others equivalences, a  relevant one  is  that  between  operator   $\mathcal L$ and equation \ref{13} which describe  the evolution inside an exponentially shaped Josephson junction. 

By means of  previous analysis,  it is also  possible to value explicitly the influences of the surface damping caused by the normal current flowing  along the junction, so  often  neglected.(see,f.i. \cite{bcs00,df213})

\vs \no Considering  $ u\, =\,e^{\,x\,\lambda /2\,}\, \bar u,    $ if  
 
 \beq
 f_1 =\, e^{\,-\,\,x\,\lambda /2\,}\, [\, \sin \,( e^{\,x\,\lambda /2\,}\,  \bar u ) \, - \gamma], 
  \eeq 
  equation (\ref{13}) became:

\begin{equation}  \label{31}
\varepsilon \bar u_{xxt}\, - \,\bar  u_{tt} \, +\, \bar u_{xx}-\, (\alpha \,+ \varepsilon \, \dfrac{\lambda^2} {4})\bar  u_t\,\,-\,  \dfrac{\lambda^2} {4}\,\bar u = \, f_1   
\end{equation}

\no   So that, removing the superscripts $ ^-  $ henceforth, and assuming:

\beq 
 a= \alpha\, + \varepsilon \, \frac{\lambda^2}{4} -\,\frac{1}{\varepsilon} \qquad \, b= \, \frac{\lambda^2}{4}\, - \frac{a}{\varepsilon}, \qquad 
 \beta \, = \dfrac{1}{\varepsilon}\eeq

  \beq   \label{32}
F\,= \, - \int_0^t\, \, e^{-\, \frac{1}{\varepsilon}(t-\tau) } f_1 (x,\tau, u) \,d \tau, \,
   \eeq  

\no  equation (\ref{31}) turns into (\ref{11}).


\begin {thebibliography}{99} 
\pdfbookmark[0]{References}{Bibliografia}



\bibitem{rb} S.Rionero {\it  Asymptotic behaviour of solutions to a nonlinear third order P.D.E. modeling physical phenomena} Boll. Unione  Matematica Italiana 2012


\bibitem {acscott}  Scott,Alwyn C. { \it The Nonlinear Universe: Chaos, Emergence, Life }.  Springer-Verlag (2007)





\bibitem {acscott02}  Scott,Alwyn C. {\it  Neuroscience A mathematical Primer }.  Springer-Verlag (2002)


\bibitem{ddf} A. D’Anna · M. De Angelis · G. Fiore {\em Existence and Uniqueness for Some 3rd Order
Dissipative Problems with Various Boundary Conditions} Acta Appl Math (2012) 122: 255-267

\bibitem{mda13} M. De Angelis, On exponentially shaped Josephson junctions Acta Appl. Math 122  iussue I (2012)
179-189


\bibitem{df13} M.D. Angelis, G. Fiore,Existence and uniqueness of solutions of a class of third order dissipative problems with various boundary conditions describing the Josephson effect, J. Math.
Anal. Appl. , 404, Issue 2, (2013) 477-490


\bibitem{df213}  M.De Angelis, G. Fiore, Diffusion effects in a superconductive model accepted in  Communications on Pure and Applied Analysis (2013)  




\bibitem{bcf}Bini D., Cherubini C., Filippi S.{\it  Viscoelastic Fizhugh-Nagumo models.} Physical Review E 041929  (2005)

\bibitem{r} Renardy, M. { \it On localized Kelvin - Voigt damping}. ZAMM Z. Angew Math Mech { 84}, (2004)


\bibitem  {dr1}De Angelis, M. Renno,P.{\it  Diffusion and wave behavior in linear Voigt model.} C. R. Mecanique {330} (2002)

\bibitem{mps}  Morro, A.,  Payne.L. E.,  Straughan,B.: { \it Decay, growth,continuous dependence and uniqueness results of generalized heat
theories}. Appl. Anal.,{ 38} (1990)


\bibitem{fr}Flavin J.N., S.Rionero{\it Qualitative estimates for Partial Differential Equations: an introduction} . Boca Raton, Florida: CRC Press(1996)


\bibitem{l}  Lamb,H.: { \it Hydrodynamics}. Cambridge University  Press  (1971)
\bibitem{dm13} M.De Angelis A priori estimates for excitable models accepted in  Meccanica  DOI...


\bibitem{de13} M.De Angelis {\em Asymptotic estimates related to an integro differential equation} accepted in Nonlinear  Dynamics and Systems Theory 


\bibitem{dr13}M.De Angelis, P. Renno  Asymptotic effects of boundary perturbations in excitable systems submitted to Discrete and Continuous Dynamical Systems - Series B  http://arxiv.org/pdf/1304.3891v1.pdf


\bibitem  {dr8}De Angelis, M. Renno,P  {\it Existence, uniqueness and a priori estimates for a non linear integro - differential equation }Ricerche di Mat. 57 (2008)

\bibitem {bp}Barone Paterno' Physical and applications of the Josephson effect 1982

 \bibitem{bcs96} A. Benabdallah; J.G.Caputo; A.C. Scott \emph{Exponentially tapered josephson flux-flow oscillator} Phy. rev. B {\bfseries{54}}, 22 16139 (1996) 

\bibitem{bcs00} A. Benabdallah; J.G.Caputo; A.C. Scott \emph{ Laminar  phase flow for an exponentially tapered josephson oscillator} J. Apl. Phys. {\bfseries{588}},6   3527 (2000)

\bibitem{cmc02} G. Carapella, N. Martucciello, and G. Costabile \emph{ Experimental investigation of flux motion in exponentially shaped Josephson junctions}PHYS REV B {\bfseries{66}}, 134531 (2002)

\bibitem{bss}T.L. Boyadjiev , E.G. Semerdjieva  Yu.M. Shukrinov    \emph{ Common features of vortex structure in long exponentially shaped Josephson junctions and Josephson junctions with inhomogeneities}
Physica C {\bfseries{460-462}} (2007) 1317-1318 (2007) 

\bibitem{j05} M. Jaworski \emph{   Exponentially  tapered Josephson junction: some analytic results }Theor and Math Phys, {\bfseries{144}}(2): 1176  1180 (2005)

\bibitem{ssb04}  Yu.M. Shukrinov, E.G. Semerdjieva and T.L. Boyadjiev \emph{ Vortex structure in exponentially shaped Josephson
junctions} J. Low Temp  Phys.  {\bfseries{19}}1/2  299 (2005)
\bibitem{j005} M. Jaworski \emph{ Fluxon dynamics in exponentially shaped Josephson
junction} Phy. rev. B {\bfseries{71}},22 (2005)       

\bibitem{cwa08} S.A. Cybart et al., \emph{ Dynes Series array of incommensurate superconducting quantum interference
devices } Appl. Phys Lett {\bfseries{93}} (2008)
\end{thebibliography}
\end{document}